# Generating VaR scenarios under Solvency II with product beta distributions

Dietmar Pfeifer[1] and Olena Ragulina[2]

October 16, 2018

**Abstract** We propose a Monte Carlo simulation method to generate stress tests by VaR scenarios under Solvency II for dependent risks on the basis of observed data. This is of particular interest for the construction of Internal Models and requirements on evaluation processes formulated in the Commission Delegated Regulation. The approach is based on former work on partition-of-unity copulas, however with a direct scenario estimation of the joint density by product beta distributions after a suitable transformation of the original data.

## 1 Introduction

The estimation of joint densities for possibly dependent random variables has a long history. Besides classical parametric methods and kernel density approaches (see e.g. SCOTT (2016)) other techniques have found interest in recent time like spline data interpolation (see e.g. SCHUMAKER (2015)). A different approach that is frequently used nowadays in insurance and finance is the decomposition of the problem into a marginal distribution estimation and the estimation of the interior dependence structure via copulas (see e.g. MCNEIL ET AL. (2015) for a general survey). In particular, Bernstein copulas and, more general, partition-of-unity copulas seem to be very well suited for Monte Carlo studies for dependent risks from which risk measures like Value at Risk (VaR) or Expected Shortfall can easily be estimated (see e.g. BLUMENTRITT (2012), CHERUBINI ET AL. (2004), COTTIN AND PFEIFER (2014), DURANTE AND SEMPI (2016), IBRAGIMOV AND PROKHOROV (2017), JOE (2015), MAI AND SCHERER (2017), MALEVERGNE AND SORNETTE (2006), RANK (2007), ROSE (2015) or SZEGÖ (2004), and for partition-of-unity copulas, in particular with applications to tail dependence, PFEIFER ET AL. (2016, 2017, 2018)). Another recent approach to tail dependence modelling via copulas is YANG ET AL. (2015). A very interesting application to claims reserving with dependence is discussed in PEŠTA AND O. OKHRIN (2014).

Reasonable VaR-estimates from original data or suitable scenarios within so-called Internal Models are of particular interest under Solvency II (see e.g. CADONI (2014), CRUZ (2009), EMBRECHTS ET AL. (2013), MAINIK (2015) or SANDSTÖM (2011)). In this paper, we propose a simple stochastic Monte Carlo algorithm beyond copulas for the generation of various VaR-scenarios that are suitable for comparison purposes in Internal Models for the calculation of solvency capital requirements. Note that the EUROPEAN UNION (2015) concerning the implementation of Solvency II in the European Union (2015) requires the consideration of such scenarios in several Articles, in particular in Article 259 on Risk Management Systems saying that insurance and reinsurance undertakings shall, where appropriate, include performance of stress tests and scenario analyses with regard to all relevant risks faced by the undertaking, in their risk-management system. The results of such analyses also have to be reported in the ORSA (Own Risk and Solvency Assessment) report as described in Article 306 of the Commission Delegated Regulation. The problem is, however, that the Commission Delegated Regulation does not make any clear statements on how such stress tests or scenario analyses have to be performed. Article 1 of the Commission Delegated Regulation defines a 'scenario analysis' as an analysis of the impact of a combination of adverse events. The Monte Carlo simulation algorithm developed in this paper allows for a mathematically rigorous description how such scenarios can be generated, being flexible enough to cover also extreme situations.

---

[1] Carl von Ossietzky Universität Oldenburg, Germany
[2] Taras Shevchenko National University of Kyiv, Ukraine

**2 The Monte Carlo algorithm**

The central idea in this paper is to transform firstly *n* marginal observations from *d* different risks with suitably estimated cumulative distribution functions (cdf's), so that the resulting data can be considered as observations from a multivariate distribution concentrated on the *d*-dimensional unit cube, similar - but typically not identical - to a copula. The next step is to approximate this distribution by a mixture of product beta distributions concentrated around each observation. This is similar to the estimation of the underlying dependence structure by a Bernstein copula or related constructions (see e.g. COTTIN AND PFEIFER (2014) and PFEIFER ET AL. (2016, 2017, 2018)). By a marginal-wise backwards transformation of the simulated multivariate distribution with the quantile functions of the originally estimated marginal cdf's we obtain realizations of an approximating distribution of the original data which allows for various VaR scenarios and VaR estimates that are particularly suitable in Internal Models under Solvency II. Note that this procedure influences the modelled dependence structure as well as the marginal distributions of the risks involved.

To be more precise, assume that $X_{ki}$ is the *i*-th observation of the *k*-th risk, for $i \in \{1,\cdots,n\}$ and $k \in \{1,\cdots,d\}$. Then, if $F_k$ denotes the true underlying cdf of the *k*-th risk then obviously $\{(F_1(X_{1i}),\cdots,F_d(X_{di}))|i=1,\cdots,n\}$ is a sample of the true underlying copula by Sklar's Theorem (cf. e.g. DURANTE AND SEMPI (2016), Chapter 2). Now if $\hat{F}_k$ denotes a suitably estimated absolutely continuous cdf for the *k*-th risk and $\hat{f}_k$ its corresponding density, define

$$\hat{h}(x_1,\cdots,x_d) := \frac{1}{n}\sum_{i=1}^{n}\prod_{k=1}^{d}b\left(x_k,(m+1)\hat{F}_k(X_{ki}),(m+1)\left(1-\hat{F}_k(X_{ki})\right)\right) \text{ for } (x_1,\cdots,x_d) \in (0,1)^d \quad (1)$$

(mixture of randomized product beta distributions) and

$$\hat{g}(y_1,\cdots,y_d) := \hat{h}\big(\hat{F}_1(y_1),\cdots,\hat{F}_d(y_d)\big) \cdot \prod_{k=1}^{d} \hat{f}_k(y_k) \text{ for } (y_1,\cdots,y_d) \in \mathbb{R}^d, \qquad (2)$$

where $b(x,\alpha,\beta) := \dfrac{x^{\alpha-1}(1-x)^{\beta-1}}{B(\alpha,\beta)}$ for $0 < x < 1$ and $\alpha, \beta > 0$ denotes the density of the Beta distribution with parameters $\alpha$ and $\beta$, and $B(\alpha,\beta) := \int_0^1 x^{\alpha-1}(1-x)^{\beta-1}\,dx$ denotes the corresponding Beta function, and $m > 0$ is a further steering parameter of the model. This approach is similar to the construction in COTTIN AND PFEIFER (2014) and resembles a classical kernel density estimate for the dependence structure where the kernel is represented by product beta densities.

Note that given $X_{ki} = z$, the expectation of the Beta distribution with parameters $(m+1)\hat{F}_k(z)$ and $(m+1)\big(1-\hat{F}_k(z)\big)$ is $\hat{F}_k(z)$ and its variance is $\dfrac{\hat{F}_k(z)\big(1-\hat{F}_k(z)\big)}{m+2} \leq \dfrac{1}{4(m+2)}$.

Seemingly $\hat{g}$ is the randomized density of a multivariate distribution (scenario distribution) that "interpolates" the original observations of the risks under investigation. This follows by similar arguments as JOE (2015), p. 8f or DURANTE AND SEMPI (2016), Remark 2.2.2 since obviously, $\hat{h}$ is the randomized density of a $d$-dimensional distribution with cdf $\hat{H}$, and $\hat{g}$ is the density of the cdf $\hat{G}$ defined by

$$\hat{G}(y_1,\cdots,y_d) := \hat{H}\big(\hat{F}_1(y_1),\cdots,\hat{F}_d(y_d)\big) \text{ for } (y_1,\cdots,y_d) \in \mathbb{R}^d. \qquad (3)$$

Note that due to the remark above, the additional parameter $m$ influences essentially the shape of the density $\hat{g}$ as the bandwidth does for kernel type density estimators. In general, we can conclude that $\hat{g}$ is more strongly concentrated around the original observations the larger $m$ is. Given the observations $X_{ki} = x_{ki}$, simulations following the cdf $\hat{G}$ or the density $\hat{g}$ can be created as follows:

1. Choose an index $I$ randomly according to a uniform distribution over $\{1,\cdots,n\}$.
2. Generate independently $d$ random variables $Z_1,\cdots,Z_d$ where $Z_k$ follows a Beta-distribution with parameters $(m+1)\hat{F}_k(x_{kI})$ and $(m+1)\big(1-\hat{F}_k(x_{kI})\big)$ (product beta distribution).
3. Set $Y_k := \hat{F}_k^{-1}(Z_k)$.

Then $(Y_1,\cdots Y_d)$ represents a Monte Carlo sample from the desired multivariate scenario distribution.

Obviously, the shape of the density $\hat{g}$ depends on $m$ as well as on the estimation of the marginal risk cdf's. Hence large sets of scenarios can be generated to estimate the VaR or other risk measures from Monte Carlo studies that embed the original data in a suitable way.

## 3 Case study

For simplicity, we will concentrate on the example data set given in COTTIN AND PFEIFER (2014) because it was also used as a data basis in several papers on partition-of-unity-copulas (PFEIFER ET AL. (2016, 2017, 2018). Here we have $d = 2$ and $n = 20$. The marginal distributions were estimated by Q-Q-plots as normal and Gumbel for the log risks, i.e. as lognormal for the first risk and Fréchet for the second risk, see the table and graphs below.

| no. | risk $X_1$ | risk $X_2$ |
|---|---|---|
| 1 | 0,468 | 0,966 |
| 2 | 9,951 | 2,679 |
| 3 | 0,866 | 0,897 |
| 4 | 6,731 | 2,249 |
| 5 | 1,421 | 0,956 |
| 6 | 2,040 | 1,141 |
| 7 | 2,967 | 1,707 |
| 8 | 1,200 | 1,008 |
| 9 | 0,426 | 1,065 |
| 10 | 1,946 | 1,162 |
| 11 | 0,676 | 0,918 |
| 12 | 1,184 | 1,336 |
| 13 | 0,960 | 0,933 |
| 14 | 1,972 | 1,077 |
| 15 | 1,549 | 1,041 |
| 16 | 0,819 | 0,899 |
| 17 | 0,063 | 0,710 |
| 18 | 1,280 | 1,118 |
| 19 | 0,824 | 0,894 |
| 20 | 0,227 | 0,837 |

Tab. 1

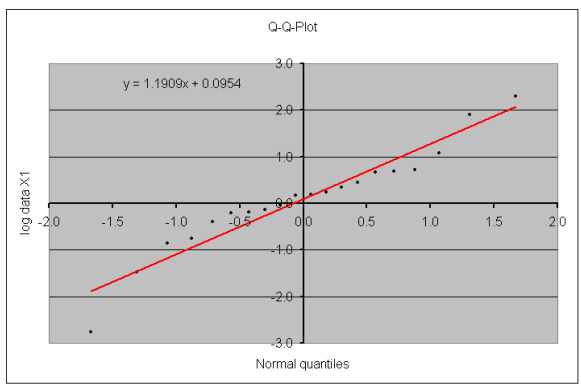 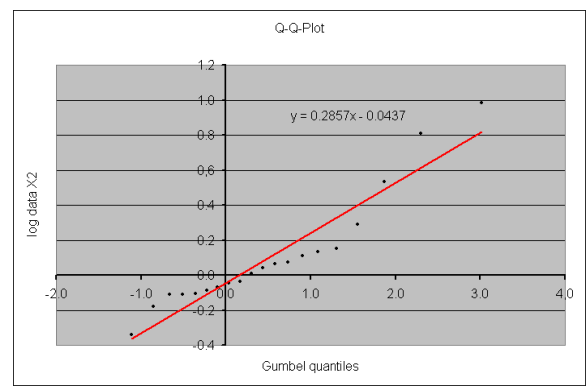

Fig. 1 — Q-Q-plot for the first risk, log data.

Fig. 2 — Q-Q-plot for the second risk, log data.

From this analysis, we get the following estimates for the location parameter $\mu$ and the scale parameter $\sigma$ for the log risks.

|          | $\mu$   | $\sigma$ |
|----------|---------|----------|
| $\ln(X_1)$ | 0.0954  | 1.1909   |
| $\ln(X_2)$ | –0.0437 | 0.2857   |

Tab. 2

The following graphs show scatterplots for various Monte Carlo simulations with the algorithm described above, for several integer values of *m*, and graphs of the contour plots of the estimated scenario density $\hat{g}$. The original data are marked by circles. The simulation size was 10,000 in each case. For comparison, we also present scatterplots for a Monte Carlo simulation with a certain adaptive kernel density estimator, where for the first risk, we use pointwise lognormal densities and for the second risk, Fréchet densities matching their modes with the data points (cf. SCOTT (2016), Chapter 6.6). In particular, the kernel density estimator used here is given by

$$\hat{f}(x,y,\sigma,\alpha) := \frac{1}{20}\sum_{i=1}^{20} k_1(x, X_{1i}, \sigma) \cdot k_2(y, X_{2i}, \alpha) \tag{4}$$

where

$$k_1(x,z,\sigma) = \frac{1}{\sqrt{2\pi}\sigma x} \exp\left(-\frac{1}{2}\left(\frac{\ln\left(\frac{x}{z}\right) - \sigma^2}{\sigma}\right)^2\right) \text{ and }$$

$$k_2(x,z,\alpha) = \frac{\alpha+1}{x}\left(\frac{z}{x}\right)^{\alpha} \exp\left(-\left(1+\frac{1}{\alpha}\right)\left(\frac{z}{x}\right)^{\alpha}\right), \ x > 0. \tag{5}$$

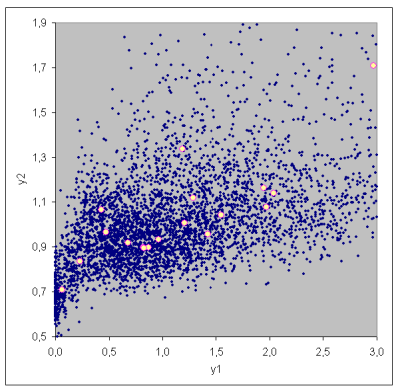 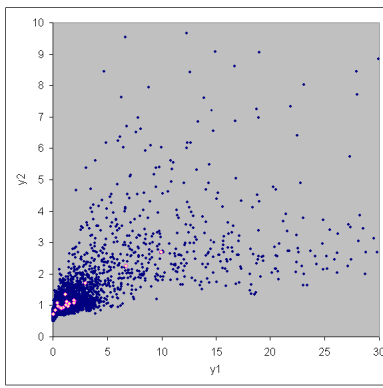 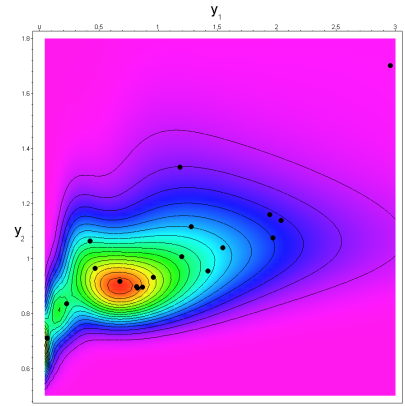

Fig. 3      Fig. 4      Fig. 5

$m = 15$; simulation scatterplot and contour plot of $\hat{g}(y_1, y_2)$

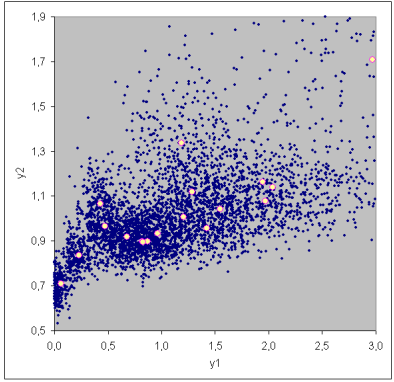 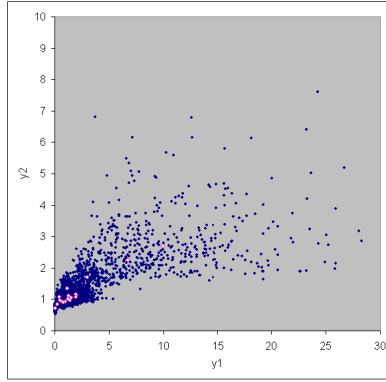 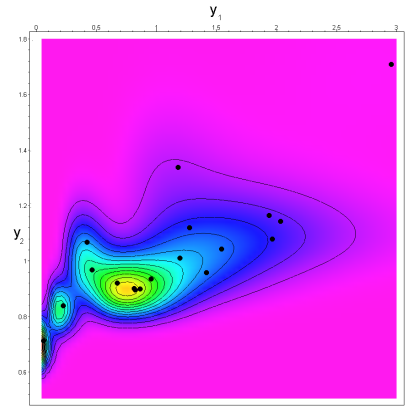

Fig. 6      Fig. 7      Fig. 8

$m = 30$; simulation scatterplot and contour plot of $\hat{g}(y_1, y_2)$

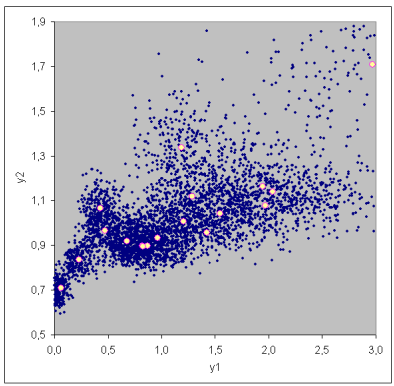 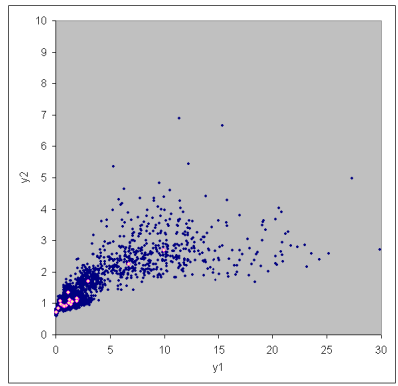 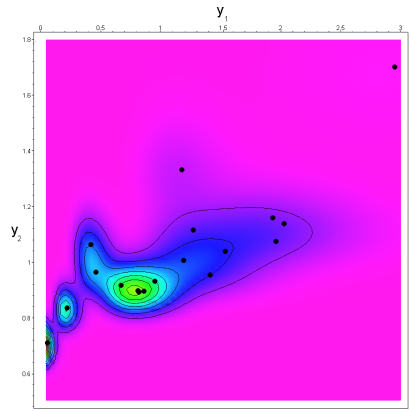

Fig. 9      Fig. 10      Fig. 11

$m = 50$; simulation scatterplot and contour plot of $\hat{g}(y_1, y_2)$

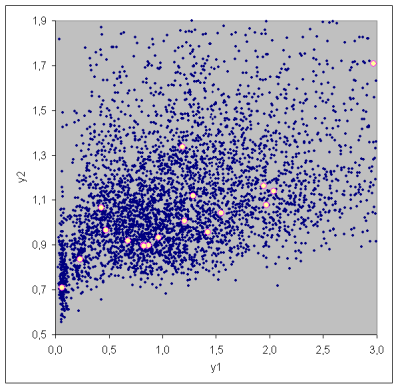 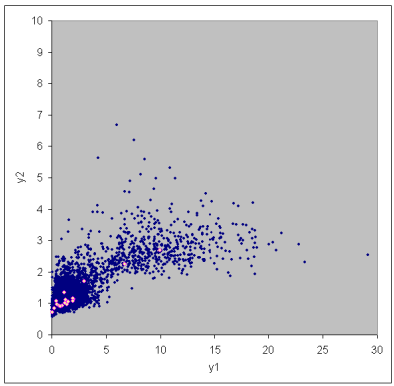 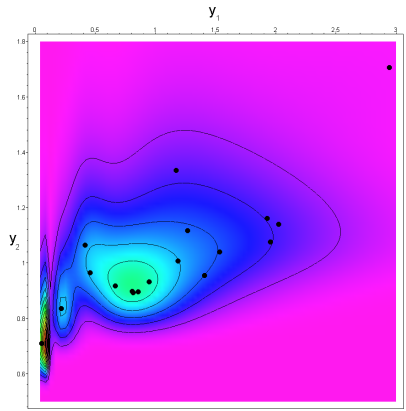

Fig. 12      Fig. 13      Fig. 14

kernel density estimate; simulation scatterplot and bivariate density contour plot

For the kernel density estimate, the parameters $\sigma = 0.3$ and $\alpha = 7$ were used.

The following table shows various estimates of the $\text{VaR}_\alpha$ for the aggregated risk ( $= 1 - \alpha$ - quantile of the sum distribution for the risks), calculated from 100,000 simulations each, where $\alpha$ denotes the risk level:

|  | $m=15$ | $m=20$ | $m=25$ | $m=30$ | $m=50$ | $m=100$ | kernel density |
|---|---|---|---|---|---|---|---|
| $\widehat{\text{VaR}}_{0.05}$ | 13.987 | 12.978 | 12.347 | 12.016 | 11.341 | 10.908 | 11.754 |
| $\widehat{\text{VaR}}_{0.01}$ | 40.637 | 31.235 | 26.989 | 23.966 | 19.498 | 16.580 | 17.272 |
| $\widehat{\text{VaR}}_{0.005}$ | 60.752 | 44.270 | 36.410 | 30.846 | 23.390 | 18.864 | 19.087 |

Tab. 3

Obviously the estimated VaR's decrease with increasing $m$ for every risk level $\alpha$, which seems reasonable since with increasing $m$, the scenario distribution is closer concentrated around the original data points, which is also clearly reflected in the graphs above. For $m \to \infty$, we would get a VaR estimate from the empirical distribution, i.e. a value of 12.630 for $\alpha \leq 0.01$ and 8.980 for $\alpha = 0.05$. Note also that with a kernel density approach, extreme scenarios can in general not be obtained.

It is interesting to observe that for $\alpha = 0.005$ (Solvency II standard) the estimated VaR is almost twice as high for $m = 15$ as in comparison to $m = 30$.

VaR estimates with a classical Bernstein copula or finite, infinite or continuous partition-of-unity copulas with or without tail dependence as in PFEIFER ET AL. (2017, 2018) typically give much smaller values. The following table lists some results for comparison. The rook copula driver for the Negative Binomial and the Gamma copula shows no tail dependence, the upper Fréchet copula (UF) driver does. For technical details, see PFEIFER ET AL. (2017, 2018).

|  | Bernstein | NB rook, $a=7$ | NB UF, $a=7$ | NB rook, $a=15$ | NB UF, $a=15$ |
|---|---|---|---|---|---|
| $\widehat{\text{VaR}}_{0.05}$ | 7.166 | 6.885 | 7.016 | 6.974 | 7.155 |
| $\widehat{\text{VaR}}_{0.01}$ | 15.634 | 15.973 | 15.744 | 15.877 | 16.059 |
| $\widehat{\text{VaR}}_{0.005}$ | 21.105 | 20.801 | 21.311 | 20.256 | 21.733 |

Tab. 4

|  | Gamma rook, $a=7$ | Gamma UF, $a=7$ | Gamma rook, $a=15$ | Gamma UF, $a=15$ |
|---|---|---|---|---|
| $\widehat{\text{VaR}}_{0.05}$ | 9.330 | 10.072 | 9.522 | 10.191 |
| $\widehat{\text{VaR}}_{0.01}$ | 18.113 | 21.224 | 18.550 | 21.428 |
| $\widehat{\text{VaR}}_{0.005}$ | 22.933 | 28.123 | 23.079 | 28.588 |

Tab. 5

The following graphs show some realizations of the induced empirical copulas (scaled rank vectors) based on 5,000 simulations for different choices of $m$ and the kernel approach outlined above. The empirical copula of the original data (scaled rank vectors) is represented by circles in each plot. For comparison purposes, we also show some realizations of the Negative Binomial (NB) and the Gamma copulas with parameters from Tab. 4 and Tab. 5, taken from PFEIFER ET AL. (2017, 2018) and COTTIN AND PFEIFER (2014).

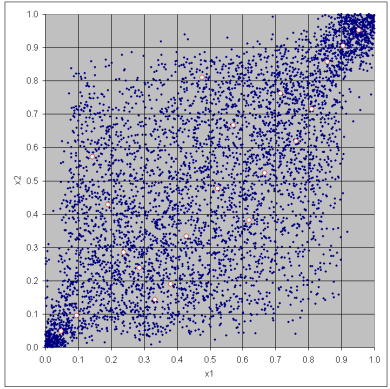
Fig. 15
$m = 15$; empirical copula

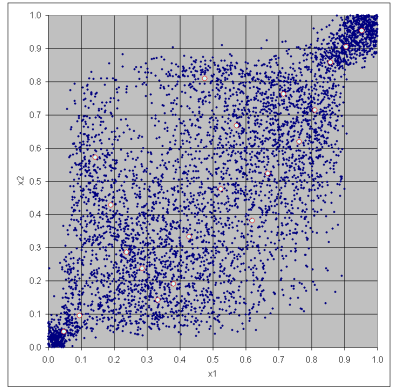
Fig. 16
$m = 30$; empirical copula

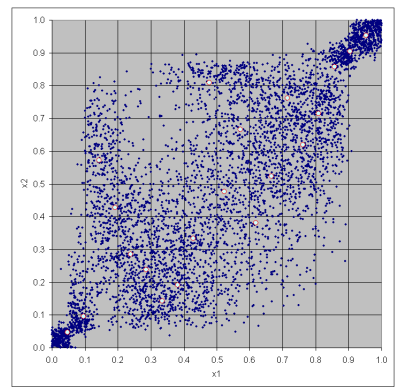
Fig. 17
$m = 50$; empirical copula

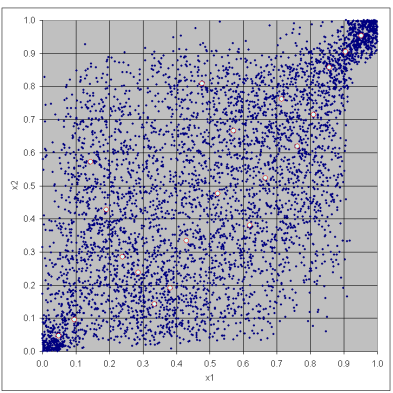
Fig. 18
empirical kernel copula

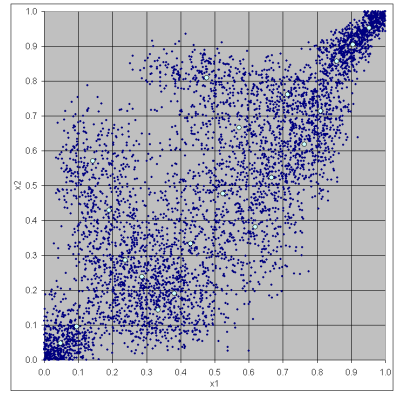
Fig. 19
$a = 15$; Gamma rook copula

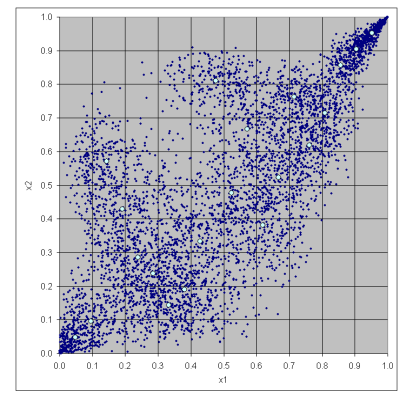
Fig. 20
$a = 15$; Gamma UF copula

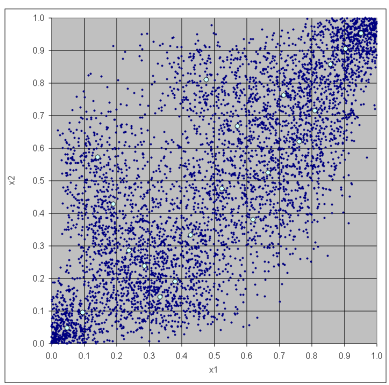
Fig. 21
$a = 15$; NB rook copula

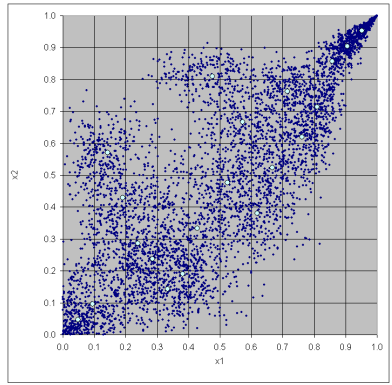
Fig. 22
$a = 15$; NB UF copula

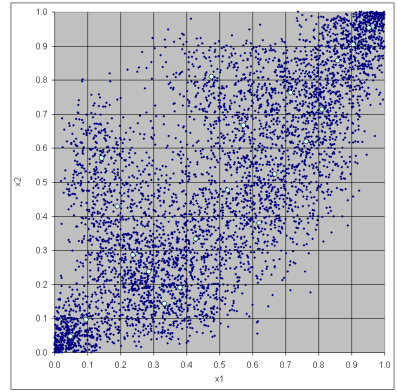
Fig. 23
Bernstein copula

Seemingly, the structure of the various copula approaches alone (with and without tail dependence) does not give any hint to the height of the VaR-estimate for the aggregate risk. An ordering of the figures according to the decreasing magnitude of the VaR-estimate for $\alpha = 0.005$ (Solvency II standard) is: Fig. 15, Fig. 16, Fig. 20, Fig. 17, Fig. 19, Fig. 22, Fig. 23, Fig. 21, and Fig. 18.

Finally, we present Q-Q-plots for the marginal distributions of the log risks from 5,000 simulations for different choices of *m* and the kernel approach outlined above. The plot positions for the theoretical quantiles are chosen with the parameters from Tab. 2. Additionally, the original data points are shown as circles.

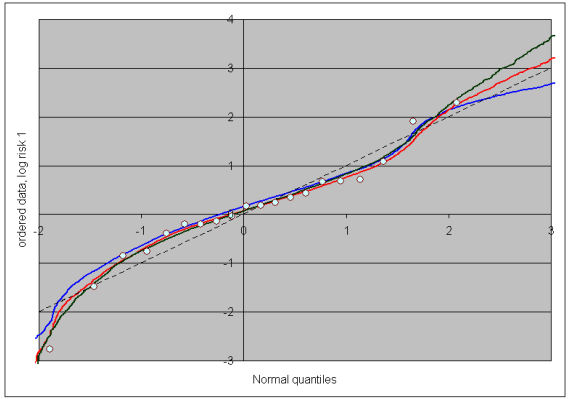

Fig. 24

Q-Q-plot for the first risk, log data.

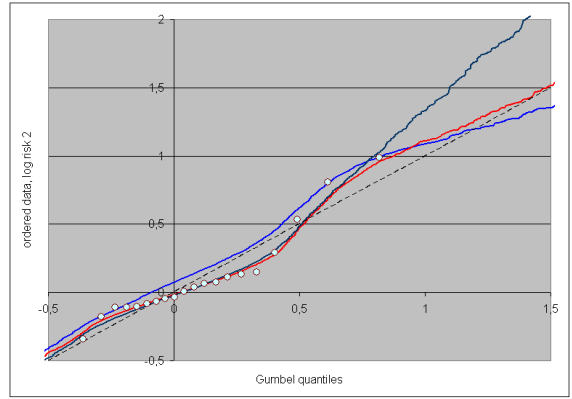

Fig. 25

Q-Q-plot for the second risk, log data.

Seemingly, the product beta and the kernel density approach are in good coincidence with the body of the data, while the product beta approach is characterized by essentially higher values in the upper tail of the marginal distributions. This emphasizes again the fact that unfavourable VaR estimates cannot be characterized by the copula structure alone but that the interplay between the dependence structure and the marginal distributions is essential, as discussed in IBRAGIMOV AND PROKHOROV (2017).

**Conclusions**

The algorithm proposed here typically generates mathematically well-defined high-score VaR-scenarios on the basis of observed losses with particular emphasis on the underlying stochastic dependence which reproduce the original data exactly, and give stress tests and scenario analyses under Solvency II a more precise meaning. It is applicable in arbitrary dimensions and generally superior to kernel density or classical and recent copula approaches, with respect to complexity, easy implementation (even in usual spreadsheet programs), and larger scenario VaR estimates. We have tested the procedure described in this paper with the 19-dimensional data set discussed in NEUMANN ET AL. (2018) and came to similar conclusions.

A crucial point here is the estimation of the marginal distributions which, of course, influences the results to a certain extend; likewise for the value of *m*. However, in any case, the original data are exactly reproduced, and the selection of the steering parameters should depend on the purpose of the application.

**Acknowledgements**

We thank the referees for a constructive criticism which lead to an improvement of the overall presentation of the content of this paper.